# Thermal Brownian Motion of Skyrmion for True Random Number Generation

Yong Yao[1,2,&], Xing Chen[1,3,&], Wang Kang[1,3,*], Youguang Zhang[1,2], Weisheng Zhao[1,3,*]

[1]Fert Beijing Research Institute, BDBC, Beihang University, Beijing 100191, China

[2]School of Electronics and Information Engineering, Beihang University, Beijing 100191, China

[3]School of Microelectronics, Beihang University, Beijing 100191, China

[&]These authors contributed equally to this work and should be considered co-first authors

*E-mails: wang.kang@buaa.edu.cn and weisheng.zhao@buaa.edu.cn

## Abstract

The true random number generators (TRNGs) have received extensive attention because of their wide applications in information transmission and encryption. The true random numbers generated by TRNG are typically applied to the encryption algorithm or security protocol of the information security core. Recently, TRNGs have also been employed in emerging stochastic/probability computing paradigm for reducing power consumption. Roughly speaking, TRNG can be divided into circuits-based, e.g., oscillator sampling or directly noise amplifying; and quantum physics-based, e.g., photoelectric effect. The former generally requires a large area and has a large power consumption, whereas the latter is intrinsic random but is more difficult to implement and usually requires additional post-processing circuitry. Very recently, magnetic skyrmion has become a promising candidate for implementing TRNG because of their nanometer size, high stability, and intrinsic thermal Brownian motion dynamics. In this work, we propose a TRNG based on continuous skyrmion thermal Brownian motion in a confined geometry at room temperature. True random bitstream can be easily obtained by periodically detecting the relative position of the skyrmion without the need for additional current pulses. More importantly, we implement a probability-adjustable TRNG, in which a desired ratio of “0” and “1” can be acquired by adding an anisotropy gradient through voltage-controlled magnetic anisotropy (VCMA) effect. The behaviors of the skyrmion-based TRNG are verified by using micromagnetic simulations. The National Institute of Standards and Technology (NIST) test results demonstrate that our proposed random number generator is TRNG with good randomness. Our research provides a new perspective for efficient TRNG realization.

## I. Introduction

Information security has become one of the most important concerns with the rapid growth of data volume by the emergence of Internet of Things and artificial intelligence. Data encryption using random numbers is an effective way to ensure the security of modern information. Some indispensable mechanisms, such as the key generation, plaintext conversion, etc., require random numbers with good statistical properties and security, which can effectively prevent unauthorized organizations from accessing data during data transfer and storage [1,2]. Furthermore, emerging computing paradigms, for example, stochastic/probability computing, rely on true random bitstream for good computing quality [3-6].

Two methods are generally used to generate the required random numbers: pseudo random number generator (PRNG) and true random number generator (TRNG) [7-9]. The bit sequence generated by the former has a certain regularity and will periodically change when the sequence length is long enough, which however, may increase the predictability of bit sequence [10]. In contrast, random numbers generated by TRNG are more secure, thus are widely used in cryptography key to protect privacy [11].

Based on the physical source of randomness, TRNG can be roughly divided into digital TRNG and analog TRNG [12,13]. The source of the entropy of the digital TRNG is mainly from a random noise source, such as the frequency of the self-excited oscillator, the metastability of the circuit components, circuit noises, and the like [10,12,14]. Due to process variation, digital TRNG faces a serious problem, that is, bit "0" and "1" in the output sequence are not evenly distributed, leading to a serious probability offset, which makes the post-processing circuit indispensable. Additional circuitry not only reduces the output bit rate, but also brings additional power and area. In addition, with the downscaling of process technology, leakage current greatly increases, resulting in even more static power consumption. Therefore, digital TRNGs built by CMOS circuits generally suffer from complicated structure, large area, and large power consumption.

The entropy source of analog TRNG is typically from intrinsic thermal noise or photoelectric effect [1,15,16]. The stochastic dynamics caused by thermal noise are naturally random and are promising for TRNG designs. Recently, a rich variety of analog TRNGs have been proposed with the advances of emerging nanoscale devices, such as memristors and magnetic tunnel junctions (MTJs) [7,14,17,18]. However, owing to their hardware structures, complex probability tracking and precise control of the amplitude and duration of the current/voltage pulses are required to produce the desired probability (50%), resulting in overhead in hardware area and power consumption. For example, the memristor-based TRNG in Ref. [17] requires a pair of SET and RESET current pulses to generate each random bit; the spin-transfer torque MTJ (STT-MTJ) based TRNG under a certain amplitude of current has inevitable probability deviation [7,19]. In Ref. [7], a parallel structure with multiple MTJs was proposed, which significantly reduces the probability deviation, but increases considerable circuit area and power consumption. Recently, a spin-orbit torque MTJ (SOT-MTJ) based TRNG was proposed in Ref. [14], but a correction circuit is needed to achieve the probability balance, also resulting in additional area and energy consumption. With the rapid research and development of emerging nanoscale devices, more efficient TRNG designs are expected.

Magnetic skyrmion is a particle-like spin configuration with prominent advantages, such as nanoscale size, high motion velocity, and low depinning current density [20-26]. These intrinsic features have been studied extensively for racetrack memory application both in theories and in experiments [27-29]. Moreover, based on experimental reports, the thermal Brownian motion of skyrmion is intrinsically random, non-repeatable and unpredictable (see Appendix A for the simulation details) [20,30-36]. These advantageous features of skyrmion make it promising for TRNG with small area and low power consumption [37-39].

In this work, we propose a TRNG design based on magnetic skyrmion. Compared to the previously mentioned TRNGs, the skyrmion-based TRNG utilizes thermal noise from temperature as the entropy source to produce an unbiased random sequence without requiring additional current pulses or compensation circuitry. Thus, the proposed TRNG has higher energy efficiency and smaller area. More importantly, we can easily configure the probability between bit "0" and bit "1" (probability adjustable) by adding an anisotropy gradient through voltage-controlled magnetic anisotropy (VCMA) effect to modulate the equilibrium position of the skyrmion [40-42]. More specifically, the perpendicular magnetic anisotropy (PMA) gradient drives skyrmion from the position of large PMA area to the small PMA area and the probability of bit "0" or "1" in the random sequence will be varied accordingly [43-45]. Our proposal is validated by micromagnetic simulations through Mumax3 with integrated modules for Brownian motion [46]. The quality of random numbers generated from our TRNG have been evaluated by the National Institute of Standards and Technology (NIST) suite. Our design provides new ideas for the source of true random data in information processing and emerging computing paradigm, e.g., stochastic computing (see Appendix C) [3-6].

## II. Results

### A. Theoretical Model

The proposed skyrmion-based TRNG consists of a ferromagnetic (FM) layer, a heavy metal (HM) layer and two MTJs [see Fig.1 (a)]. A skyrmion is nucleated at the center of the FM layer and performs random motion. These two MTJs are used to detect whether the skyrmion appears on the left or right side of the FM layer, respectively [47]. The skyrmion motion within the chamber is affected by thermal disturbance [48]. The magnetization dynamics of the skyrmion can be modeled by a modified Landau-Lifshitz-Gilbert equation:

$$\frac{\partial \vec{m}}{\partial t} = -\gamma \vec{m} \times \vec{H}_{\text{eff}} + \alpha \vec{m} \times \frac{\partial \vec{m}}{\partial t} \tag{1}$$

where the effective field $\vec{H}_{\text{eff}}$ includes the demagnetization field ($\vec{H}_{\text{demag}}$), the interface perpendicular anisotropy field ($\vec{H}_{\text{ani}}$), the exchange field ($\vec{H}_{\text{ex}}$), the Dzyaloshinskii-Moriya interaction (DMI) field ($\vec{H}_{\text{DMI}}$), and a stochastic field ($\vec{H}_{\text{thermal}}$) due to thermal disturbance. Here, $\vec{m}$ is the unit magnetization vector of the region, $\alpha$ is the Gilbert damping constant, $\gamma$ is the gyromagnetic ratio.

In our system, the simulation region is an approximately elliptical chamber consisting of a rectangular region of 101 nm × 74 nm and two semicircular regions with a radius of 37 nm, as show in Fig. 1(b). Magnetization dynamics of the skyrmion can be strongly influenced by the random thermal fluctuations. In practical situations, the output bits are assigned depending on the differential voltage $\Delta V$ $(\Delta V = V_{\text{left}} - V_{\text{right}})$ of the two MTJs, as shown in Fig. 1(c). Specifically, if $\Delta V$ is positive, we regard it as a bit of “0”, corresponding to the presence of the skyrmion on the left; otherwise, a bit “1” is indicated, or vice versa.

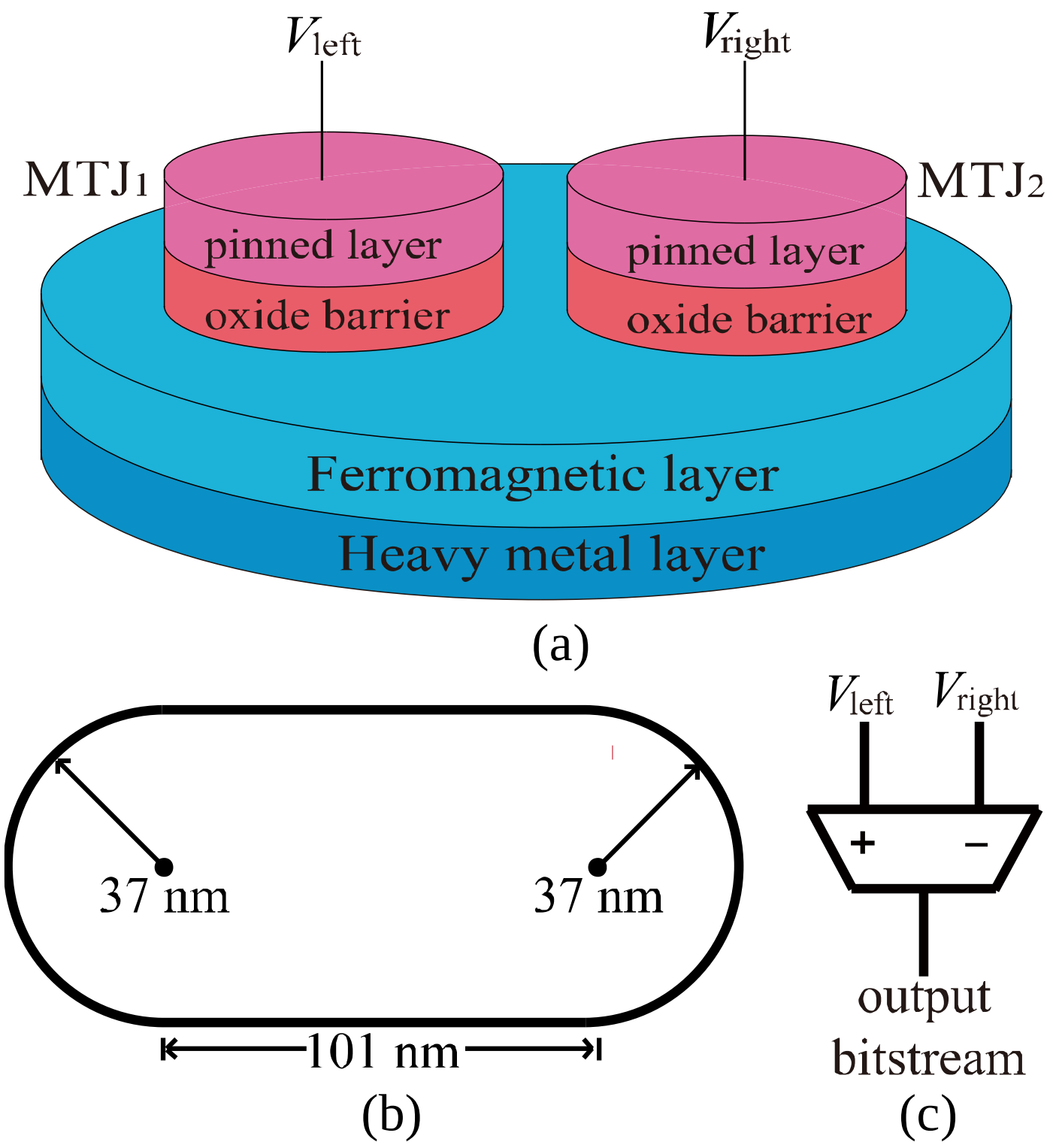

Fig. 1 (a) The schematic structure of the skyrmion-based TRNG: the ferromagnetic layer is used for skyrmion Brownian motion and the two MTJs are used to detect the skyrmion. (b) The top view of the FM chamber: a rectangular region of 101 nm × 74 nm and two semicircular regions with a radius of 37 nm. (c) The schematic of the comparator. It is assigned a bit of “0” if the skyrmion is on the left; otherwise, a bit of “1” is indicated.

The skyrmion Brownian motion has been studied in [20,30-36]. For a non-confined two-dimensional system, thermal diffusion of skyrmion can be simply expressed as follows [31],

$$\langle [r_{x,y}(t+t^*) - r_{x,y}(t)]^2 \rangle = \langle (\Delta r_{x,y})^2 \rangle = 4\mathfrak{D}_{dc} t^* \quad (3)$$

Here, $r_{x,y}$ is the position of the skyrmion center $(x, y)$, $t^*$ is the time interval between two selected data points. The left side of the equation expresses the mean squared displacement (MSD) of a skyrmion. The diffusion constant $\mathfrak{D}_{dc}$ is established as,

$$\mathfrak{D}_{dc} = k_B T \frac{\alpha \mathcal{D}}{\mathcal{G}^2 + (\alpha \mathcal{D})^2} \quad (4)$$

where $\mathcal{G} \propto 4\pi Q$ is the gyrocoupling strength that is related to the skyrmion topological number $(Q)$, $\mathcal{D}$ is the dissipative factor that depends on the skyrmion profile [20,49]. In the two-dimensional limit, the topological number is $Q = -\frac{1}{4\pi} \int \vec{m} \cdot \left( \frac{\partial \vec{m}}{\partial x} \times \frac{\partial \vec{m}}{\partial y} \right) dxdy$, where $\vec{m} = \vec{M} / M_s$ is the reduced magnetization vector and $M_S$ is the saturation magnetization [20]. Eq. (3) indicates that $\mathfrak{D}_{dc}$ can be evaluated from the linear matching of the MSD as a function of $t^*$. Meanwhile, Eq. (4) reveals a linear dependence of $\mathfrak{D}_{dc}$ on T. Therefore, we can measure $\mathfrak{D}_{dc}$ based on the linear fitting of the MSD with regard to T and $t^*$. Fig. 2(a) shows that the MSD is linearly related to the time interval $t^*$ when $t^*$ is smaller than 2 ns. However, as $t^*$ increases, the simulation results [see Fig. 2(b)] show that the MSD no longer linearly increases as $t^*$ increases due to the boundary effect. More specifically, as $t^*$ increases, the distance of skyrmion motion increases while the displacement may decrease because the skyrmion is constrained inside the chamber, which also indicates that the MSD will not grow indefinitely with the increment of time.

In the linear region, it is difficult for a skyrmion to move from one side to the other at a small interval. Accordingly, a fairly long continuous “0” or “1” will appear in the sequence, and its randomness will be very poor. In the nonlinear region, a small MSD may represent a large moving distance. That is, skyrmion may move from one side of the chamber to the other, and then move back (see Appendix B).

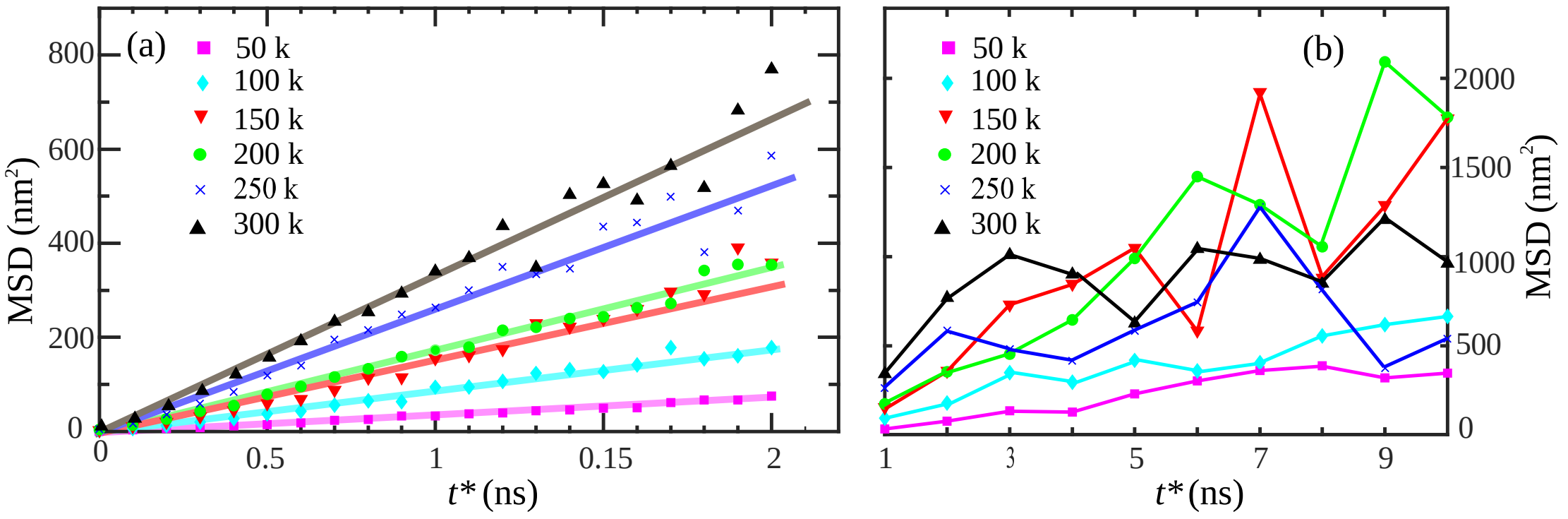

Fig. 2 (a) The mean squared displacements (MSDs) and their linear fits at selected temperatures with small intervals $t^*$. (b) Nonlinear relation between the MSDs and $t^*$ at large intervals. The MSDs still increases as $t^*$ increases and there is a weaker linearity between the MSDs and $t^*$ at lower temperatures. The MSDs and $t^*$ are completely nonlinear dependency at higher temperatures.

Similarly, Fig. 3(a) displays that the MSD is also linearly dependent on the temperature in the case of small intervals, which is consistent with Eq. (4). Here, the thermal effect induced by temperature plays a dominant role on the displacement of the skyrmion. In contrast, the boundary has a greater impact on the skyrmion if $t^*$ is larger than 3 ns, leading to the nonlinearity [see Fig. 3(b)]. To obtain an output sequence with good randomness, a relatively large MSD is necessary. Accordingly, high temperature and long interval should be considered, but the frequency of random sequence will therefore decrease. In our simulations, we detect the position of skyrmion every 10 ns at room temperature (300 K) considering both the frequency and randomness of the sequence. Higher frequencies can be obtained by raising the temperature, changing the shape of the chamber, or material parameters, etc.

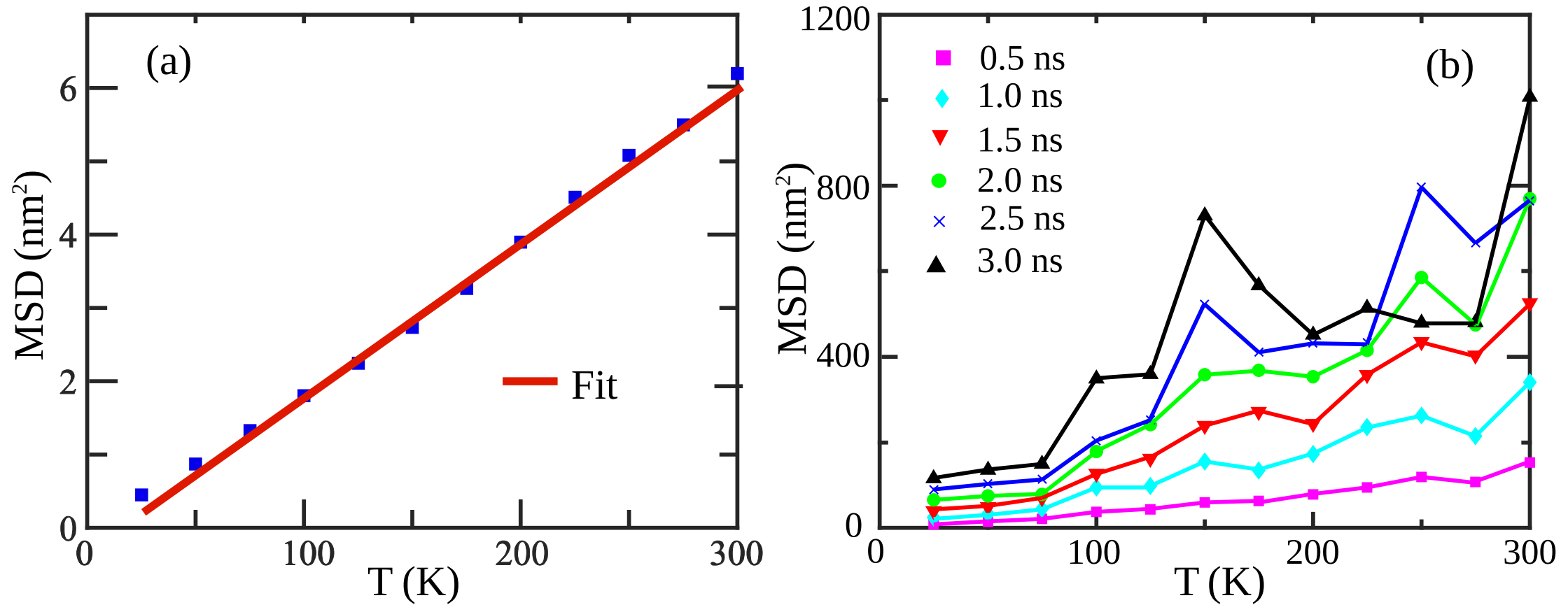

Fig.3 The MSD as a function of temperature $T$ for skyrmion with a step of (a) 20 ps, and (b) relatively large intervals.

According to the model mentioned above, we adopt the following material parameters in our simulations: the exchange stiffness $A = 15\ \mathrm{pJ/m}$, saturation magnetization $M_s = 580\ \mathrm{kA/m}$, Landau-Lifshitz damping constant $\alpha = 0.3$, interface-induce DMI constant $D_{\mathrm{ind}} = 3\ \mathrm{mJ/m^2}$, and default PMA constant of the FM layer $K_u = 0.8\ \mathrm{MJ/m^3}$. Besides, the default mesh size of 1 nm × 1 nm × 1 nm is used in our simulations.

### B. TRNG with equally distributed bit “0” and “1”

As discussed above, an isolated skyrmion will randomly diffuse inside the chamber under the combined effect of the temperature and the boundary. Random sequence can then be obtained by locating the position of the skyrmion. When the center of the skyrmion is on the left side of the chamber, it is defined as bit “0”; otherwise it is regarded as bit “1”.

Fig. 4(a) shows the trajectory of the skyrmion within 5.7 μs (570 stages). According to the simulation results, the skyrmion is located 286 times on the left side of the chamber and 284 times on the right side of the chamber. The frequency of occurrence on both sides is almost the same. This indicates that our proposed TRNG can provide unbiased output, which is the premise of examining the quality of random numbers. Fig. 4(b) shows the position of the skyrmion in the *x* axis in order to clearly observe the arrangement of “0” and “1” in the sequence. Fig. 4 (c) presents 50 ideal outputs selected from our random sequence.

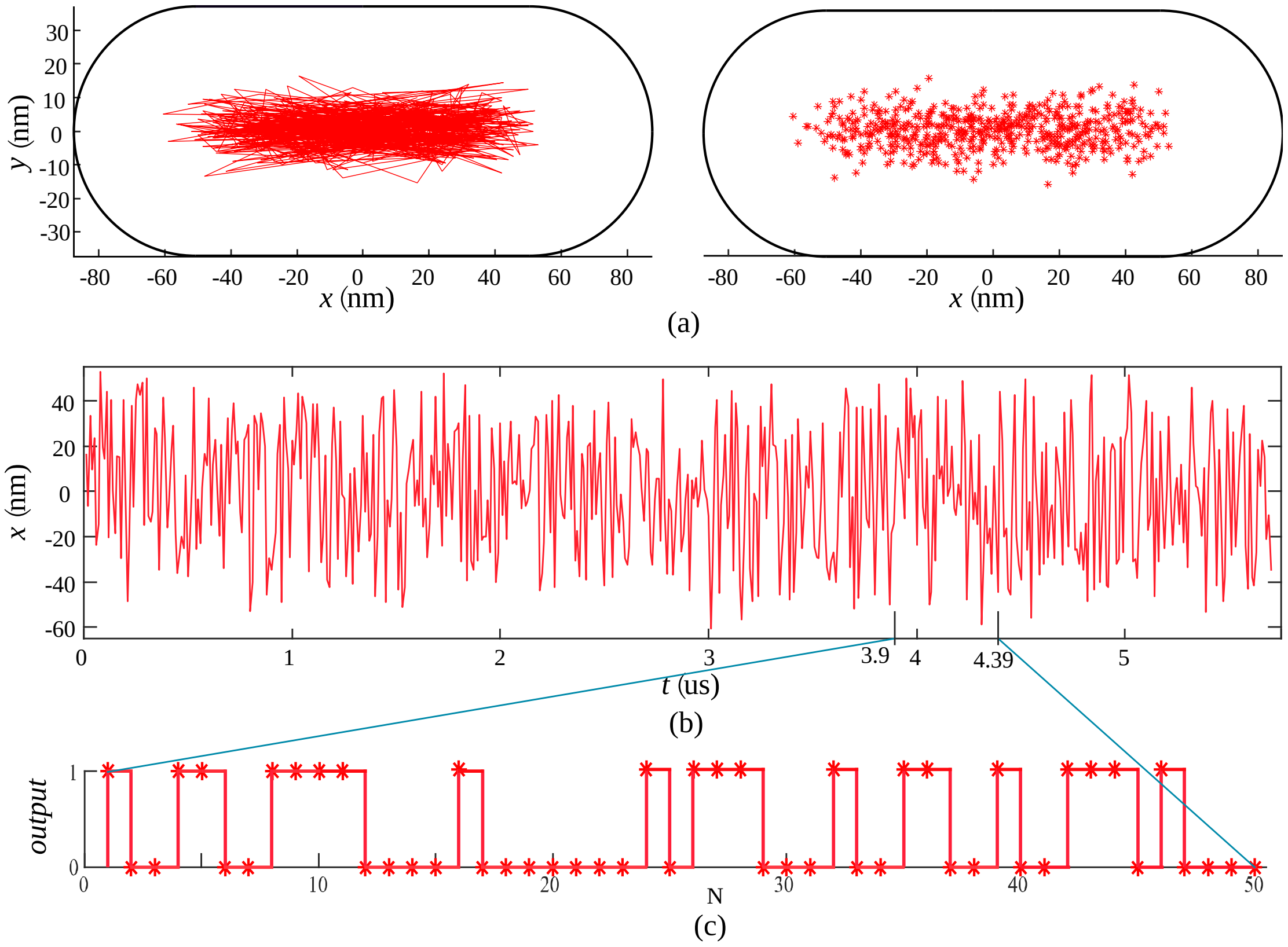

Fig. 4 Random sequence generated by skyrmion Brownian motion. (a) The motion trajectory of the skyrmion in the chamber. (b) A random sequence of length 580

obtained by detecting the position of skyrmion in the $x$-axis direction. (c) Theoretical outputs from 3.9 $\mu s$ to 4.39 $\mu s$.

Further, the quality of our random numbers is evaluated by the NIST suite. The NIST suite is a statistical package which focuses on the non-randomness possibly existing in the tested data [50]. The test can only be performed normally if the "0" and "1" in the ASCII sequence are evenly distributed, that is, the unbiased sequence is a prerequisite for ensuring the success of the test. The NIST test results of our TRNG are shown in Table I, where 9 tests are considered. The value of P calculated during every test is to evaluate the randomness under this test and a large P indicates a better randomness. If P is greater than 0.01, the test is successfully passed. Correspondingly, the binary sequence can be considered as true random number sequence, and the proposed RNG is TRNG. Table I demonstrates that the our skyrmion based random number generator (RNG) is a TRNG with fairly good randomness.

Table. I NIST test results of the random numbers

| | Test | P-value | | Success/Fail |
|---|---|---|---|---|
| 1 | Frequency | 0.716998 | | Success |
| 2 | Block Frequency | 0.119433 | | Success |
| 3 | Runs | 0.337499 | | Success |
| 4 | Longest Run of Ones | 0.808069 | | Success |
| 5 | FFT | 0.305078 | | Success |
| 6 | Approximate | 1.000000 | | Success |
| 7 | Cumulative Sums | 0.604709 | 0.329312 | Success |
| 8 | Serial | 0.824844 | 0.907266 | Success |
| 9 | Non-Overlapping Template Matching | All sub-test success | | |

### C. Probability adjustable TRNG

The non-uniform distribution of "0" or "1" bitstream can be implemented with peripheral circuits, such as cascading XOR gates [2,10,12]. However, such peripheral circuits typically consume a large amount of energy and occupy additional hardware areas. In contrast, our proposed TRNG can generate unbalanced output bitstream simply by a voltage gradient. The voltage gradient changes the value of PMA ($K_{\mathrm{u}}$) of the material, thus producing an anisotropy gradient, which drives skyrmion from the high PMA area to the lower counterpart. Regardless of the influence of temperature, the equilibrium position of the skyrmion is the center of chamber. With the combination of the gradient and the boundary, the equilibrium of the skyrmion moves a distance (*d*) from the center to the low PMA region, depending on the PMA difference $\Delta K_{\mathrm{u}} = K_{\mathrm{uv}} - K_{\mathrm{u0}}$, where $K_{\mathrm{uv}}$ is the maximum value, and $K_{\mathrm{u0}}$ is the minimum value of material [see Fig. 5(a)]. Considering the thermal disturbance, the skyrmion moves around the new equilibrium position, so the probability of skyrmion appearing on the left and right sides of the area will be different. It can be seen from Fig. 5(b) that the equilibrium position of the skyrmion is constantly shifted to the left with the increase of $K_{\mathrm{uv}}$. The uneven distribution of skyrmion position for 1000 intervals ($t^*$= 1 ns) is displayed in Fig. 5(c, d, e, f), in which the left side is much denser than the other side with the increase of $K_{\mathrm{uv}}$.

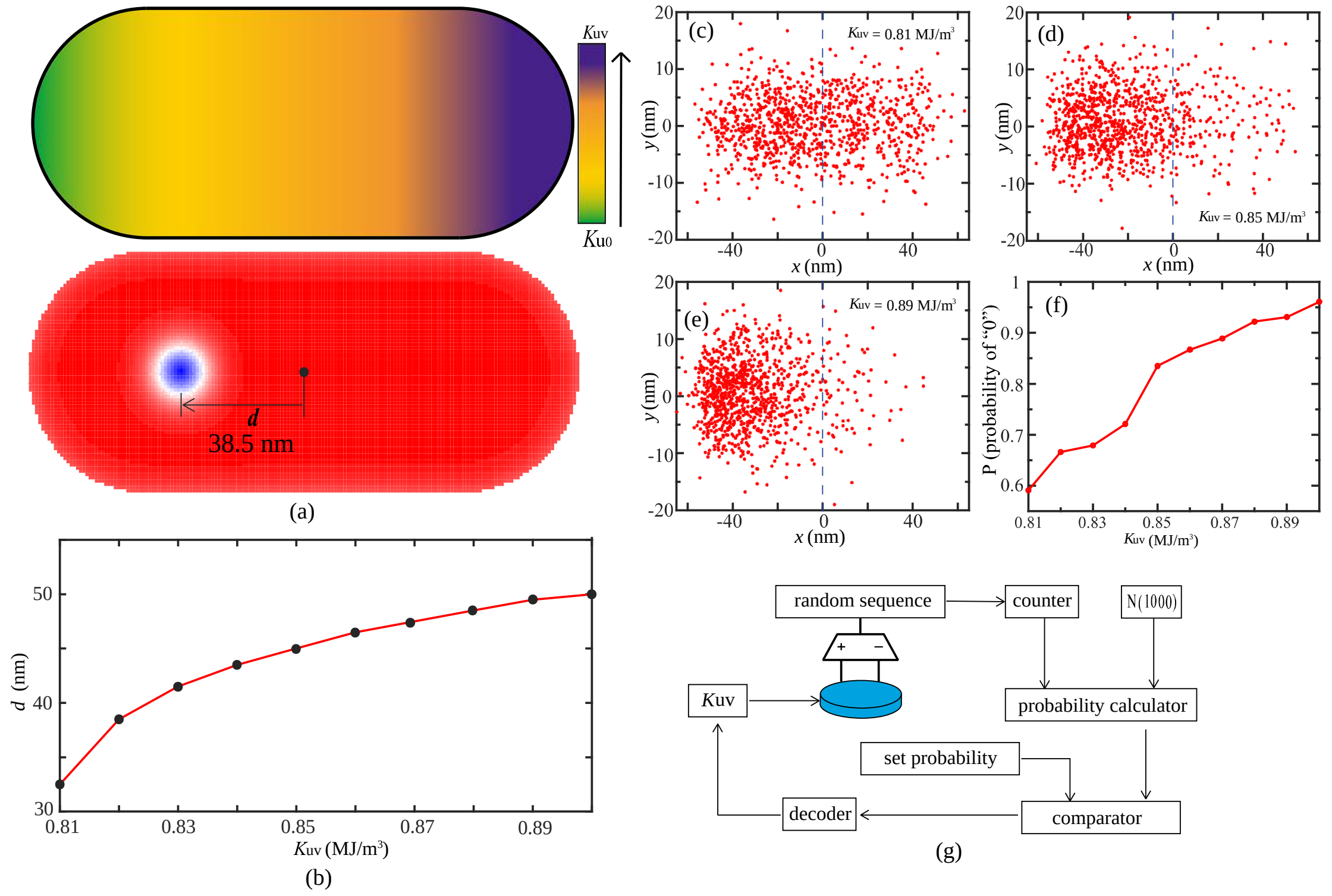

Fig. 5 (a) The schematic of the PMA gradient when $K_{uv} = 0.82\ \mathrm{MJ/m^3}$. The equilibrium position of skyrmion moves a distance of $d = 38.5\ \mathrm{nm}$. (b) The equilibrium position of the skyrmion as a function of $K_{uv}$. (c, d, e) The skyrmion position distribution when $K_{uv} = 0.81\ \mathrm{MJ/m^3}$, $K_{uv} = 0.85\ \mathrm{MJ/m^3}$, and $K_{uv} = 0.89\ \mathrm{MJ/m^3}$, respectively. (f) The probability of "0" in the output bitstream as a function of $K_{uv}$. (g) The flow chart used to precisely control the distribution of the output bitstream.

The probability of "0" or "1" in the output bitstream can be precisely controlled by a negative feedback circuit which includes a counter, a probability calculator, a comparator and a decoder, as shown in Fig. 5(g). The counter is used to count the number of "0" in a random sequence of length N obtained from the skyrmion based TRNG. The probability of "0" is obtained by a division operation in the probability calculator. The comparator transmits a signal which is the result from comparing the calculated probability and the set probability to the decoder. The core of the decoder is a CMOS transistor array which converts the signal into a voltage generating the PMA gradient. A larger voltage is supplied when the calculated probability is smaller than the set probability to generate a larger gradient, thus the calculated probability is increased.

## Conclusions

We implement a skyrmion based TRNG utilizing the thermal induced skyrmion Brownian motion property. Random number sequence with a 50% distribution can be obtained, without additional excitations, which is energy efficient compared to other TRNGs. The NIST test results indicate that the generated random numbers have fairly good randomness. Furthermore, our proposed TRNG can be adjusted to produce an output sequence with the desired probability of "0" and "1" using the anisotropy gradient. This work provides a new perspective to implement efficient TRNG for information processing and non-von Neuman computing paradigms.

## APPENDLX A: UNPREDICTABILITY AND NON-REPRODUCIBILITY OF SKYRMION BROWNIAN MOTION

The random numbers generated by a TRNG must be unpredictable and non-repeatable. Therefore, the dynamics of skyrmion must be unpredictable and non-repeatable as well to make sure the skyrmion based RNG is a TRNG. The interaction between the skyrmion and the boundary is determined whenever it is located in the chamber. However, random fluctuations caused by temperature change the direction or speed of the skyrmion. Consequently, the subsequent trajectory of the skyrmion is varied. In other words, two identical skyrmions have various dynamics under thermal fluctuations, thus the numbers represented by the position of skyrmion should also be random. As shown in Fig. 6 (a), we select a skyrmion profile near the center as the initial state, and there are two different motion trajectories. Within the first 30 ns, the distance between these two skyrmions is very small. In the next 20 ns, they move further apart. Then at $t$ = 50 ns, these two skyrmions are located in the left half side and the right half side, respectively, representing different bit values. The same initial condition, however, yields different results, which effectively demonstrate that the dynamics of skyrmions under temperature disturbances are unpredictable. Hence, it is feasible to use the intrinsic thermal induced skyrmion motion to implement TRNG.

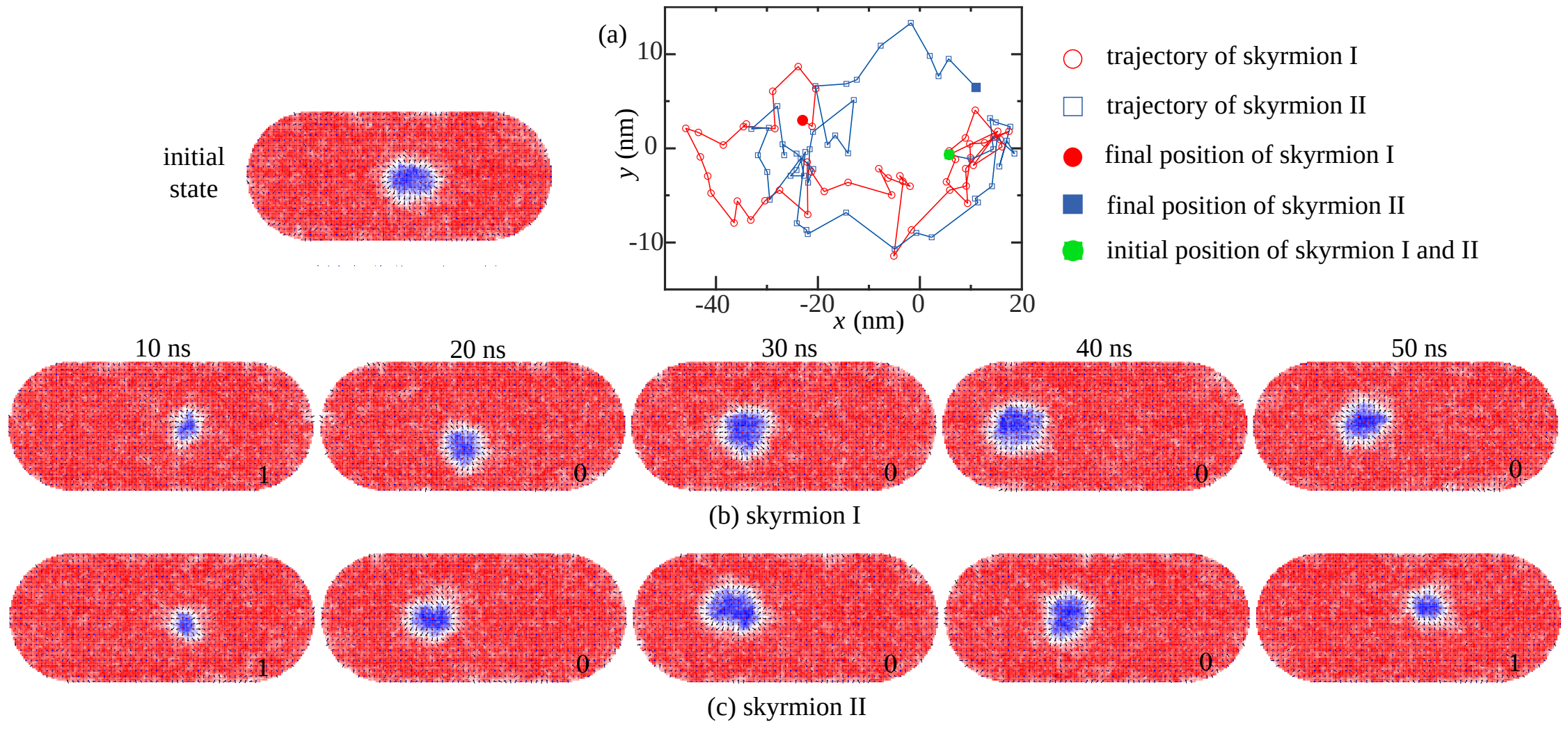

Fig. 6 (a) Different trajectories of two skyrmions with identical initial condition. (b), (c) Random numbers represented by skyrmion I and II at different times, respectively.

## APPENDLX B: RANDOMNESS OF THE SKYRMION

In order to have a good randomness, it is undesirable that there are a large number of continuous "0" or "1" in our random sequence. In the case where $t^* = 10$ ns, the skyrmion is constantly moving within the area during two adjacent detections. Thus, there will be the presence of bit "1" between two adjacent bits "0" if $t^*$ is shortened to 2 ns. This process can be observed in Fig. 7, where the skyrmion is initially on the left side of the chamber, then moves to the right side of the chamber, and finally moves back to the left side of the chamber. We can shorten the interval while keeping randomness by raising the temperature.

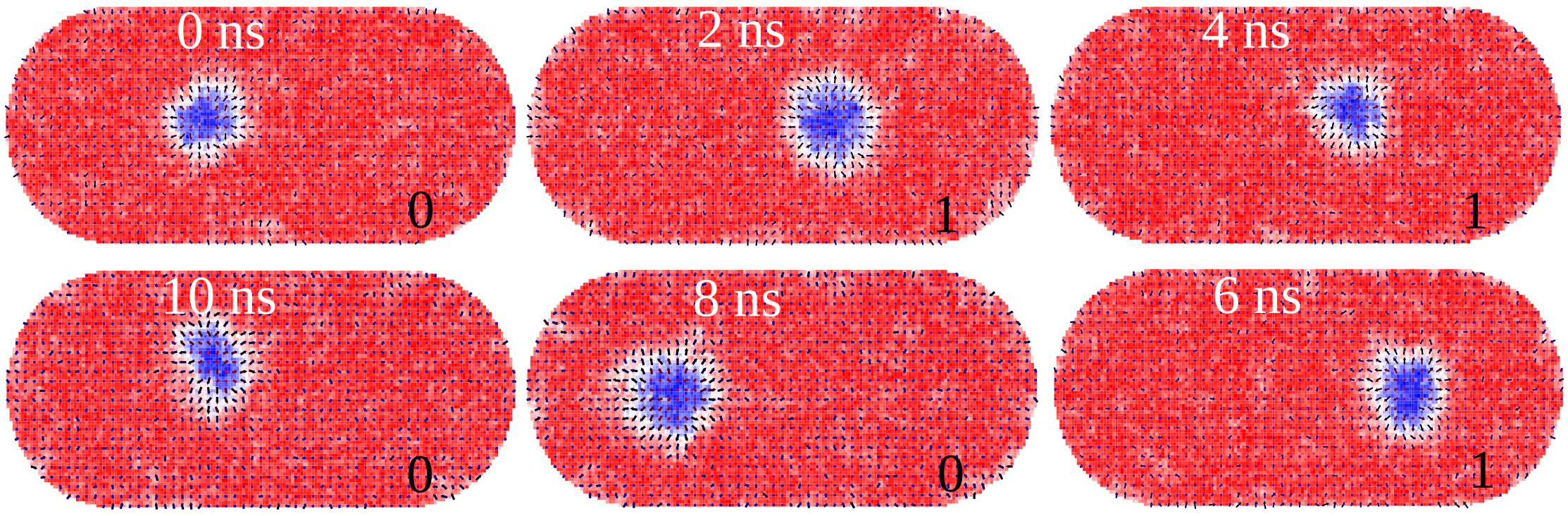

Fig. 7 There is a bit “1” between two adjacent bits “0” if the selected time interval is shortened from 10 ns to 2 ns. The skyrmion moves from the left area to the right side of the area, then goes back to the left side of the area between two adjacent output bits.

## APPENDLX C: STOCHASTIC COMPUTNG

The development of mainstream computing hardware is now limited by some practical application conditions: area cost, power consumption and reliability. At the same time, industrial manufacturing process at nanoscale node can easily lead to data errors. Emerging stochastic computing (SC) paradigm has recently received wide attention due to its unique fault tolerance [3-6]. An essential feature of SC is that the bitstream should be true random. According to the Bernoulli’s law of large numbers, the probability that each bit being “1” can be represented by the proportion of the number of bits “1” in the bitstream. In general, the implementation of a multiplication operation requires combination of many logic gates, which requires complex logic transformations. However, multiplication operation can be achieved by an AND gate in SC. As shown in Fig. 8(a), in the traditional two-level expression, the data representation could be greatly changed if there is a bit state transition in the output bitstream. In contrast, the result of random calculation will only change from 5/16 to 4/16 or 6/16, which extremely increases the fault tolerance. There is a problem with SCs, that is, there is a certain deviation between the operation result and the expected result. But the accuracy of SC grows with the length *n* of the bitstream, as shown in Fig. 8(b). We select some random numbers generated by our proposed TRNG and use them to make random operations with different data lengths. As the length of data increases, the error rate of random operations is significantly reduced.

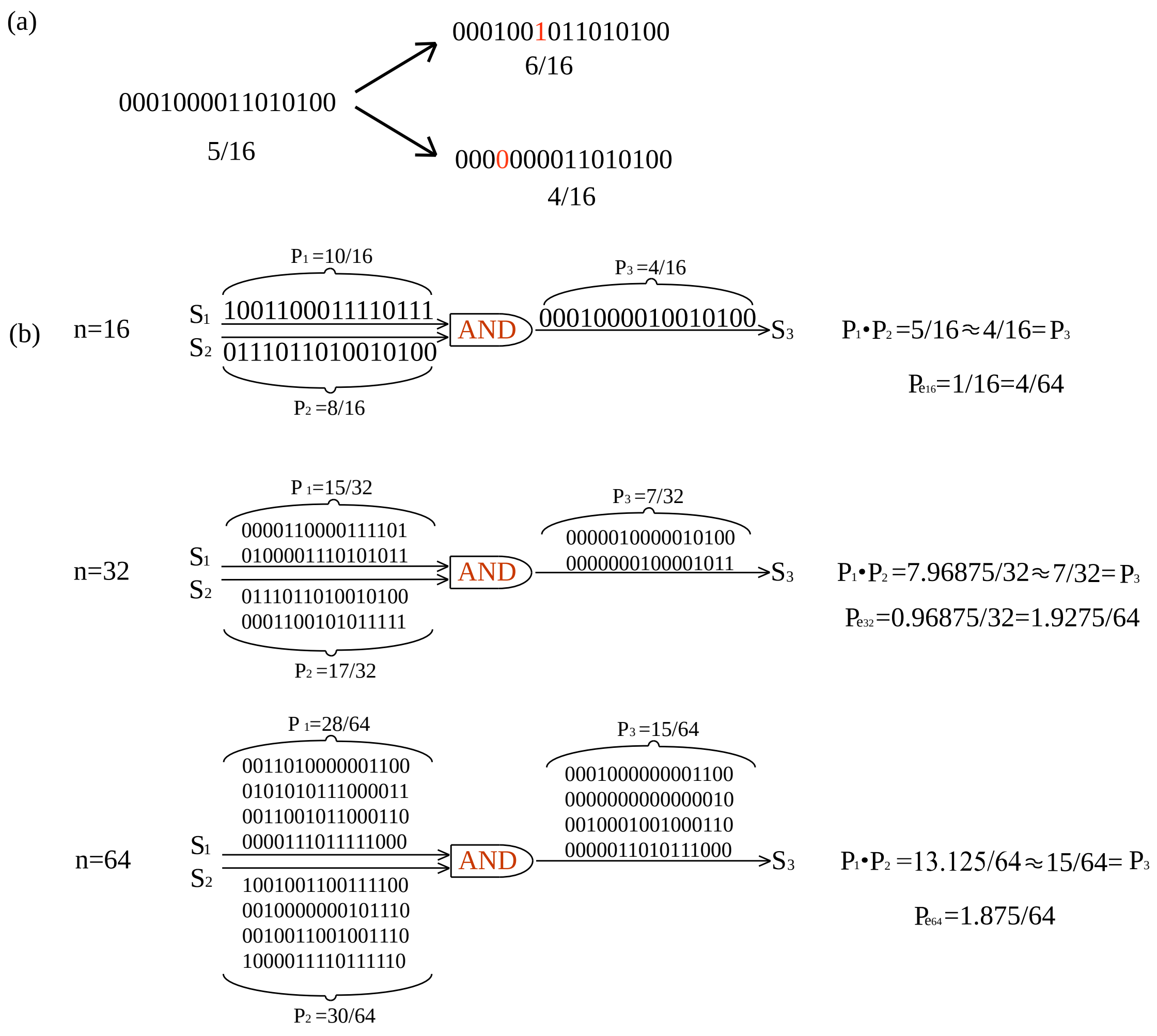

Fig. 8 (a) Small variation of data in random operations. (b) The accuracy of stochastic computing under different data lengths at n = 16, n = 32, and n = 64 respectively.